# Revealing the long-range coupling for multi-dimensional metasurface multiplexer

Ouling Wu[1,2], Chao Qian[1,2,*], Guangfeng You[1,2], and Hongsheng Chen[1,2,*]

[1] *ZJU-UIUC Institute, Interdisciplinary Center for Quantum Information, State Key Laboratory of Extreme Photonics and Instrumentation, Zhejiang University, Hangzhou 310027, China.*

[2] *ZJU-Hangzhou Global Science and Technology Innovation Center, Zhejiang Key Laboratory of Intelligent Electromagnetic Control and Advanced Electronic Integration, Zhejiang University, Hangzhou 310027, China.*

*Corresponding author: chaoq@intl.zju.edu.cn (C. Qian); hansomchen@zju.edu.cn (H. Chen)*

**Abstract:** Metasurface coupling constitutes a fundamental yet intricate electromagnetic interaction that occurs within a lattice of artificial subwavelength unit cells. Despite its prevalence, such coupling is typically ignored in conventional metasurface design frameworks due to the high characterization complexity, leading to suboptimal device performance. Here, we reveal a distinctive long-range coupling that exceeds an order of magnitude compared with the interaction range of evanescent waves, substantially enriching the metasurface design landscapes. This coupling exhibits pronounced graph topological features, and we design a graph neural network (GNN) to accurately abstract its inherent physics. Through strategic enhancement of the coupling effects, the discrete metasurface responses are transformed into continuous states, thereby unlocking diverse multiplexing channels. By further integrating the GNN into an inverse design agent, we tailor the multi-channel global response of metasurface to support simultaneous multiplexing across angle, frequency, and polarization domains. Experimentally, we demonstrate a compact metasurface multiplexer with eight independent channels, showcasing its potential for next-generation vehicular networks. This work establishes a new paradigm for highly integrated multifunctional metasurfaces, with promising prospects for high-density optical storage, information encryption, and high-capacity wireless communication.

**Introduction**

Metasurface coupling describes an elusive interaction among a lattice of artificially engineered unit cells. While such coupling can be leveraged for a variety of advanced applications, such as enhancing light-matter interactions[1-3] and converting confined waveguide modes into free-space propagating waves[4-7], its pronounced nonlinearity and complexity pose significant challenges to accurate modeling and deterministic design for user-defined functionalities. For example, in the inverse design of large-scale metasurfaces, the coupling undermines rapid scattering-field calculation methods that rely on local periodic approximations (LPA), severely degrading the device performance[8-11]. Moreover, for multifunctional metasurfaces, such as those supporting orbital angular momentum (OAM) multiplexing, coupling inevitably introduces crosstalk among different channels[12-16]. Therefore, considerable efforts have been devoted to mitigate the coupling effects. Grounded in the conviction that the coupling exhibits an extremely limited subwavelength interaction range, a conventional strategy is to increase the inter-unit-cell spacing[17-23]. While this approach can effectively reduce coupling strength, it comes at the cost of limited integration density and compromised spatial resolution. Another widely employed method involves inserting metallic shielding frames around individual unit cells[24-29]. Despite effectively isolating adjacent units and minimizing coupling, this technology induces substantial dispersion, thereby restricting the operating bandwidth. Ultimately, existing methods for suppressing coupling effects invariably impose performance trade-offs, limiting their practicality in high-performance metasurface design.

An alternative paradigm in metasurface design is to strategically exploit the coupling effects rather than suppress them, potentially unlocking unprecedented functionalities and enhanced performance. Several studies have been made to incorporate coupling effects into the metasurface design. One representative approach is to employ theoretical models for characterizing coupling effects and guiding the design process[30-34]. Although existing models such as coupled-mode theory (CMT) have proven effective in weak coupling scenarios, they often struggle to capture strongly complex coupling behaviors. Another widely adopted approach relies on supercell design[35-39], where different unit cells are grouped into a single meta-atom, and their collective response enabled by local coupling is optimized via theoretical analysis or full-wave simulation. This method supports multifunctional device design but remains limited to short-range interactions within individual supercells, neglecting inter-supercell coupling. Consequently, existing design frameworks are largely confined to weak and localized coupling, leaving the novel physics and potential application of extensive coupling in metasurfaces a promising field for exploration.

In this work, we explicitly reveal the long-range coupling effects within the metasurface using an ingeniously designed GNN framework to mimic the topological network among unit cells. By strategically enhancing and controlling the coupling effects, we enable a structurally simple metasurface to support multi-dimensional multiplexing capabilities absent in the coupling-free regime. Through GNN analysis, we demonstrate that the long-range coupling collectively reshapes the global metasurface response, transforming the originally discrete phase states into a continuous 2π phase coverage. This coupling-mediated response exhibits high sensitivity to incident wave properties, substantially expanding the design freedom of metasurfaces. Integrated with the inverse design pipeline, the GNN facilitates the realization of a compact metasurface multiplexer supporting eight independent channels across frequency, angle, and polarization domains. This highly integrated device shows great potential for enhancing vehicular networks, enabling higher data throughput and improved communication efficiency between vehicles and infrastructure[40, 41]. Our study establishes a paradigm for multifunctional metasurface design based on global collective coupling effects rather than local unit-cell engineering, opening avenues for highly integrated electromagnetic devices.

**Results**

**Physical analysis of coupling effects**

When illuminated by incident waves, complex coupling arises within metasurfaces as a result of the interaction among near-fields and the overlap of electromagnetic modes[42, 43]. This coupling induces pronounced non-local effects in individual unit cells, making their responses deviate from those obtained under periodic boundary conditions. Furthermore, we find that the incident wave can exert a distinct influence on the metasurface response that has been altered by the coupling effects, as illustrated in Fig. 1. This phenomenon can be specifically analyzed from the perspective of coupled-mode theory[44]. The coupled-mode equations for metasurfaces can be written as follows

$$\begin{cases} \dfrac{d\boldsymbol{a}}{dt} = \left(j\boldsymbol{\Omega} - \dfrac{\mathbf{D}^\dagger \mathbf{D}}{2}\right)\boldsymbol{a} + \mathbf{D}^T \mathbf{S}_{\text{in}} \\ \mathbf{S}_{\text{out}} = \mathbf{C}\mathbf{S}_{\text{in}} + \mathbf{D}\boldsymbol{a} \end{cases} \tag{1}$$

where $\boldsymbol{a}$ represents the mode amplitude and $\mathbf{D}$ represents the coupling coefficient matrix. The diagonal elements of the matrix $\boldsymbol{\Omega}$ represents the resonant frequencies of each unit cell, while the non-diagonal elements represent the near-field coupling among unit cells. The matrix $\mathbf{C}$ describes the reflection and transmission coefficients of the background. In the steady state, $\mathbf{S}_{\text{out}}$ for the working frequency $\omega_0$ can be obtained by substituting $\boldsymbol{a} = \boldsymbol{a}e^{j\omega_0 t}$, $\mathbf{S}_{\text{in}} = \mathbf{S}_{\text{in}} e^{j\omega_0 t}$, $\mathbf{S}_{\text{out}} = \mathbf{S}_{\text{out}} e^{j\omega_0 t}$ into Eq. (1).

$$\mathbf{S}_{\text{out}} = \left(\mathbf{C} + \mathbf{D}\left(j\omega_0 - j\boldsymbol{\Omega} + \dfrac{\mathbf{D}^\dagger \mathbf{D}}{2}\right)^{-1} \mathbf{D}^T\right)\mathbf{S}_{\text{in}} \tag{2}$$

Now we can clearly observe that the transfer function between the incident wave $\mathbf{S}_{\text{in}}$ and the outgoing wave $\mathbf{S}_{\text{out}}$ is directly influenced by the near-field coupling effects $\boldsymbol{\Omega}$, which is relevant to the electromagnetic mode supported by the unit cells. Consider the fundamental transverse electric (TE) modes, the near-field coupling can be approximated as

$$\Omega_{mm'} = \dfrac{1}{2}\sqrt{\dfrac{\omega_m \omega_{m'}}{Q_m Q_{m'}}} Y_0(k_0 |r_m - r_{m'}|) \tag{3}$$

where $m$ and $m'$ represent two coupled unit cells and $r$ represent their positions. $Y_0$ represents the second kind Bessel function and $k_0$ is the wave number. The two critical parameters $\omega$ and $Q$ represent the resonant frequency and quality factor of the unit cells respectively. Given that previous research has revealed that the incidence angle can induce a shift in the resonant frequency of the unit cell[45], according to Eq. (3), we can find that the incidence angle influences the coupling effects by altering the resonant frequencies of the unit cells and ultimately changing the response of the metasurface. Similarly, altering the operating frequency can also influence the coupling effects. Moreover, it is noteworthy that the coupling demonstrated in Eq. (3) is based on TE modes, and the incident waves with different polarizations can induce various modes within the unit cells, thus altering their near-field coupling[46]. Therefore, we can conclude that the characteristics of the incident wave, including the incidence angle, frequency, and polarization, can influence the intricate interactions within the metasurface, thereby changing its overall response.

**Observation of the long-range coupling in metasurfaces**

While the above-mentioned theoretical analysis has demonstrated that coupling effects are able to change the response of metasurfaces, in scenarios with weak coupling, these changes are so slight and short-range that they cannot be effectively utilized in practice. To enhance the coupling effects, we have designed a compact metasurface composed of four different types of unit cells, including round, cross, Jerusalem cross and square (Supplementary information 1). Owing to the close arrangement and pronounced boundary asymmetry, significant interaction is induced among the unit cells. To clearly demonstrate the response changes resulting from enhanced coupling effects, we first analyze the characteristics of unit cells under traditional periodic boundary conditions at the operating frequency of 11 GHz. As shown in Fig. 2b, all the unit cells exhibit conspicuous angle-insensitive characteristics. Under 0° and 45° incidences, their phase responses remain almost unchanged and all of them display high transmittance. Moreover, owing to the $C_4$ structural symmetry, the unit cell exhibits identical phase responses under both ±45° incidences and both X- and Y-polarized waves.

However, in the presence of strong and long-range coupling, these characteristics undergo a fundamental transformation, presenting a distinct situation (Supplementary information 2). To capture the intricate coupling effects precisely and comprehensively, we have carefully designed a GNN framework. As shown in Fig. 2a, this framework consists of two key modules, the graph convolution module and the U-Net module. The coupling effects among unit cells are governed by their topological network, which allows the metasurface to be modeled by a graph. In this graph, each unit cell is represented as a node, while the adjacency relationships between them define edges. Meanwhile, the structural parameters and positional information of the unit cells are assigned as corresponding node features and edge features. The graph will first be fed into the graph convolution module composed of a reduce function and an apply function. The reduce function calculates the coupling effects $c$ from surrounding unit cells leveraging both node features $n$ and edge features $e$. These computed effects are then additively aggregated and passed together with the original node features to the apply function. Based on these, the apply function computes the updated node features $n'$, which represent the unit cell responses that have incorporated coupling effects. The above process can be summarized as follows:

$$n_i' = f_a\left(n_i, \sum f_r(n_i, n_j, e_{ij})\right) \quad (4)$$

where $f_r$ and $f_a$ represent the reduce function and apply function respectively. Eq. (4) reveals the physical logic of the graph convolution module in processing coupling effects, indicating the high interpretability of the GNN. With the coupling physics abstracted by this module, the U-net subsequently analyzes and fuses multi-scale features to obtain the global response of the metasurface. The GNN is trained on full-wave simulation data and achieves precise prediction of phase/amplitude responses for coupled metasurfaces (Supplementary information 3). Furthermore, it also scales to arbitrary-large metasurfaces with ultra-fast inference speed. Even for a 200 × 200 metasurface that is extremely time-consuming for full-wave simulation, the GNN can accomplish the prediction within sub-second time, significantly facilitating the rapid inverse design of large-scale metasurfaces (Supplementary information 4).

Figure 2c presents the GNN-predicted phase of a randomly generated 26 × 26 metasurface under 0° incidence. Unlike the discrete four-level phase under periodic boundary conditions, the phase here exhibits a continuous distribution due to the coupling. To further illustrate the impact of coupling effects on metasurface response, we have replaced the four central unit cells of the metasurface with their adjacent ones respectively, and the resulting near-field intensity variations at half a wavelength are presented in Fig. 2d. It can be observed that the intensity variations extend over a distance exceeding five wavelengths, which is ten times the coupling length of the evanescent wave. These sophisticated long-range coupling effects render the local response of the metasurface highly dependent on the collective behavior of surrounding unit cells, thus manifesting extraordinary properties. We have also analyzed the phase distributions under different incidence angles, as illustrated in Fig. 2e. In both 0° and 45° cases, the phase responses fully cover the range of 2π, confirming the feasibility of achieving continuous phase modulation with discrete unit cells. Moreover, the significant difference between the two incidence angles demonstrates that the incidence angle strongly influences the response of a metasurface with long-range coupling — a finding consistent with the conclusion derived from Eq. (3).

**Implementation of multi-dimensional multiplexing**

In physical analysis, we have revealed that the characteristics of incident waves, including the incidence angle, frequency, and polarization, can influence the coupling within the metasurface and thus alter its global response. The precise characterization of coupling effects by GNN further makes it possible to achieve functional multiplexing in the metasurface. Since the metasurface comprises four different types of unit cells, we have devised a dedicated discrete optimization algorithm and integrated the GNN into the algorithmic workflow to efficiently achieve the inverse design targets (Methods). As an example, we have designed an angle-multiplexed holographic metasurface with a size of 36 × 36. As illustrated in Fig. 3a, when illuminated at angles of 0°, 45°, and -45°, the metasurface projects three distinct holographic patterns of "Z", "J", and "U" respectively onto a plane 0.1 meters away. During the inverse design process, the GNN predicts the phase and amplitude responses of the metasurface under three incidence angles respectively, which are then transformed into near-fields on the target plane through Rayleigh-Sommerfeld diffraction theory (Methods). Crucially, if coupling effects are neglected, the metasurface exhibits identical discrete phase responses across all angles and fails to function as intended. The design results in Figs. 3b and 3c show strong agreement between the GNN prediction and full-wave simulation results, validating the effectiveness of our angular multiplexing scheme.

The exotic behaviors induced by long-range coupling are not confined to the near-field of metasurface but extend to the far-field, and we will elaborate on this point by implementing abnormal far-field customization. To efficiently derive the precise far-field pattern of the metasurface, we employ the plane wave expansion method to convert the near-field obtained through the GNN into the far-field (Methods). We first implement the polarization multiplexing, where the designed metasurface deflects the beam by -30° under X-polarized wave illumination and by 30° under Y-polarized wave illumination, as depicted in Fig. 3e. The GNN-predicted phase responses reveal that this functionality arises from opposite phase gradients under the two polarization states. Notably, without considering coupling effects, all $C_4$-symmetric unit cells would produce identical far-field responses for both polarizations, making this polarization-sensitive behavior impossible. Meanwhile, in the scenario of frequency multiplexing shown in Fig. 3f, when the designed metasurface is exposed to a plane wave at 11 GHz ($f_1$), its phase response exhibits gradient changes towards two opposite directions, thereby generating two beam pointing directions. In contrast, when it is illuminated at 12 GHz ($f_2$), the nearly uniform phase response produces a pencil-shaped beam pointing at 0°.

Furthermore, the long-range coupling endows metasurfaces with the capability of simultaneous multiplexing across multiple dimensions. As shown in Fig. 4a, variations in the incident wave's angle, frequency, and polarization elicit distinct responses from the metasurface, which subsequently projects four sequential holographic numerals onto a target plane 0.1 m away via diffraction. Figures 4b and 4c present the GNN-predicted results and full-wave simulation results of the designed metasurface respectively. The Pearson correlation coefficients between them all exceed 0.8, convincingly demonstrating the accuracy and effectiveness of the multi-dimensional multiplexing metasurface. To comprehensively evaluate the inter-channel crosstalk, we have calculated the confusion matrices for the ideal, GNN-predicted, and simulation outcomes, as shown in Fig. 4e. Although inherent shape similarities between digits "1" and "4", as well as "2" and "3", lead to relatively high correlations even in the ideal case, digits with distinct shapes exhibit consistently low correlation coefficients across all three scenarios. This confirms minimal crosstalk between different multiplexing channels.

**8-channel metasurface multiplexer for vehicular networks**

long-range coupling empowers metasurfaces to achieve multi-dimensional multiplexing, substantially expanding their capabilities for manipulating electromagnetic waves. By simultaneously adjusting the incidence angle, polarization, and frequency of the incident waves, each with two distinct states, the metasurface can generate eight physical channels. This multi-channel characteristic makes the metasurface particularly appropriate for wireless communications, localization, and integrated sensing. Figure 5a illustrates the application scenario of the metasurface for the vehicular network. Due to its compact and lightweight properties, the metasurface can be seamlessly integrated into vehicles. As a receiving device, it separates incoming wave signals from vehicles, road infrastructure, and other environmental sources based on their angular, polarization, and spectral characteristics. This capability facilitates critical functions such as position identification and traffic flow perception. Furthermore, by enabling physical-layer signal separation, the metasurface effectively enhances communication capacity and ensures high-speed transmission for large-volume data.

In the experiment, we designed and fabricated a compact 36 × 36 metasurface multiplexer using standard PCB manufacturing process. As illustrated in Fig. 5c, this device focuses incident waves with distinct characteristics at different positions on the focal plane. The experimental setup in a microwave anechoic chamber is shown in Fig. 5b. A wide-band horn antenna, connected to a vector network analyzer (VNA), illuminates the metasurface. The polarization and incidence angle of the incident waves are controlled by adjusting the horn's orientation. The electric field on the focal plane (located 0.1 m from the metasurface) is measured by a probe, which is also connected to the VNA. The detected power is derived from the $S_{21}$ parameter. As shown in Fig. 5d, the experimental results demonstrate pronounced focusing in the target areas under all incident conditions, while the power remains low in non-target regions. This provides compelling evidence that our metasurface multiplexer can effectively manipulate waves in distinct channels based on their inherent properties, highlighting its strong potential for practical applications.

**Conclusion**

In summary, we have exploited the long-range coupling effects among unit cells to tailor the global response of metasurfaces, thus enabling multi-dimensional multiplexing capabilities. A carefully devised GNN accurately characterizes these coupling effects, and its integration with a discrete optimization algorithm has facilitated the design and experimental validation of a compact metasurface that simultaneously multiplexes incidence angle, polarization, and frequency. Remarkably, the structure is composed solely of conventional unit cells, highlighting the universality of coupling effects and their generalizability across configurations. More importantly, by enhancing the extensive coupling effects among discrete unit cells, we have successfully achieved continuously adjustable responses. This outcome substantiates the feasibility of devising high-performance devices using coupled discrete unit cells. The discrete design approach will substantially simplify the manufacturing process of large-scale metasurfaces while improving their manufacturing robustness, particularly in terahertz and optical regimes. Our work establishes a new paradigm for the design and miniaturization of multifunctional devices, holding great promise in high-capacity communication, information storage, integrated sensing, and beyond.

# Methods

## Discrete optimization algorithm

We employ the discrete optimization algorithm to devise the layout of unit cells in order to achieve the desired near-field or far-field characteristics of the metasurface. For a metasurface comprising N × N unit cells, we randomly generate its initial arrangement matrix **U** first. Subsequently, we calculate the near-field or far-field of the metasurface utilizing the GNN and obtain the loss value $l$ between it and the target. The loss function is defined as the mean squared error (MSE) loss. Then, we replace the first unit cell with one of the four types of units and calculate the new loss value $l'$ resulting from the candidate arrangement **U**′. If $l' < l$, we accept the **U**′ and proceed to update the next unit cell. After N × N iterations, the metasurface undergoes a complete update of its arrangement. Multiple such rounds are executed until the loss converges, yielding the final, optimized metasurface arrangement. The specific workflow is as follows.

| **Algorithm 1:** Discrete optimization algorithm | |
|---|---|
| **Input:** | Iteration number M |
| **Output:** | Arrangement matrix **U** |
| 1. | Randomly initialize arrangement matrix **U** |
| 2. | Calculate the loss value $l$ utilizing the GNN |
| 3. | **For** *m* = 1, M **do** |
| 4. |     **For** *i* = 1, N × N **do** |
| 5. |         $u_i \leftarrow u_i'$, get the candidate arrangement **U**′ |
| 6. |         Calculate the new loss value $l'$ according to **U**′ |
| 7. |         If $l' < l$ : accept the candidate arrangement, **U** ← **U**′ |
| 8. |         If *i* < N × N: *i* ← *i* + 1; Otherwise: *m* ← *m* + 1, *i* ← 1 |
| 9. |     **End for** |
| 10. | **End for** |

## Rayleigh-Sommerfeld diffraction theory

The Rayleigh-Sommerfeld diffraction is usually utilized for calculating the propagation of electromagnetic waves in free space, and the process of it can be mathematically expressed as

$$\mathrm{E}(x,y,z) = \iint \frac{z}{r^2}\left(\frac{1}{2\pi r} + \frac{1}{j\lambda}\right) e^{-j\frac{2\pi r}{\lambda}} \mathrm{E}(x',y',0) dx' dy' \qquad (5)$$

where $\mathrm{E}(x',y',0)$ and $\mathrm{E}(x,y,z)$ represent the electric field at a point on the source plane and the target plane at a distance $z$ from the source plane. The variable $r$ denotes the distance between the point $(x,y,z)$ on the target plane and the center point of the source plane, which can be expressed as $r = \sqrt{x^2 + y^2 + z^2}$ .

## Plane wave expansion method

The electric field of a monochromatic wave radiated by an aperture can be written as a superposition of plane waves of the form $\boldsymbol{f}(k_x, k_y) \exp(-j\boldsymbol{kr})$. The function $\boldsymbol{f}(k_x, k_y)$ is the vector amplitude of the wave, and $k_x$ and $k_y$ are the spectral frequencies which extend over the entire frequency spectrum. Thus, the electric field in spherical coordinates can be written as

$$\mathbf{E}(r,\theta,\varphi) \cong j\frac{ke^{-jkr}}{2\pi r}\left[\boldsymbol{\theta}(f_x\cos\varphi + f_y\sin\varphi) + \boldsymbol{\varphi}\cos\theta(-f_x\sin\varphi + f_y\cos\varphi)\right] \quad (6)$$

where $f_x$ and $f_y$ are the components of $\boldsymbol{f}$ in the $x$ and $y$ directions respectively. As $r$ tends to infinity, the far-field of the radiating aperture can be obtained.

**Acknowledgments**

The work at Zhejiang University was sponsored by the National Natural Science Foundation of China (NNSFC) under Grant Nos. 62101485, 62422514, 62471432, and 92564301, the Key Research and Development Program of Zhejiang Province under Grant No. 2022C01036, and the Fundamental Research Funds for the Central Universities.

**Author contributions**

O.W. and C.Q. conceived the idea of this research. O.W. performed the theoretical analysis and simulations. O.W. and C.Q. wrote the paper. All authors shared their insights and contributed to discussions on the results. C.Q. and H.C. supervised the project.

**Data availability**

The data that support the results of this study are available from the authors upon reasonable request.

**Competing interests**

The authors declare no competing financial interests.

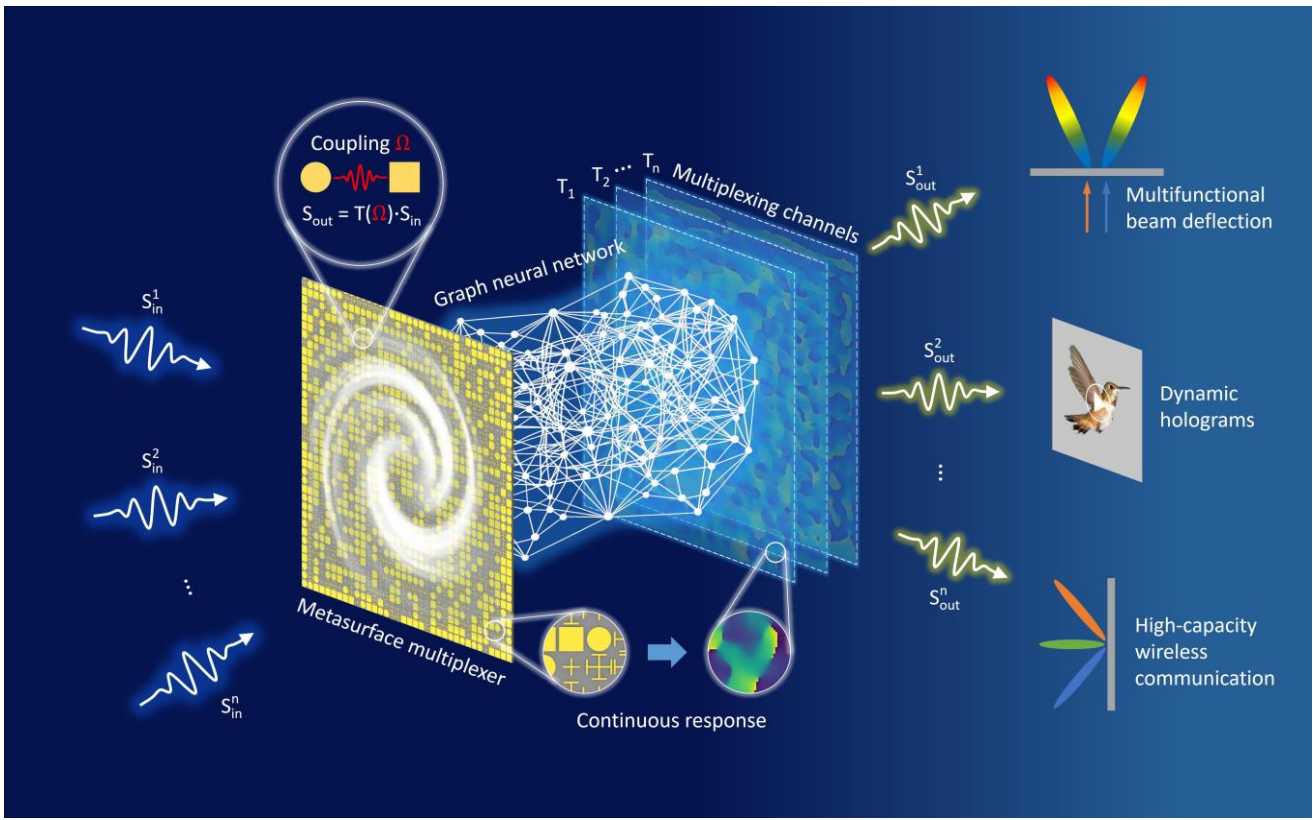

**Fig. 1 | Schematic of the long-range coupling for metasurface multiplexer.** The sophisticated and ubiquitous electromagnetic coupling in metasurface holds great potential for wave manipulation. By strategically enhancing the long-range inter-unit-cell interactions, the global response of metasurface becomes continuous and is sensitive to the incident waves, enabling diverse multiplexing channels and an expanded design space. The graph neural network can precisely capture these interactions based on topological correlations among unit cells, facilitating the effective harnessing of coupling effects. Consequently, by strategically tailoring long-range coupling, highly compact multidimensional metasurface multiplexers can be designed, offering significant promise for applications in dynamic holography and high-capacity wireless communications.

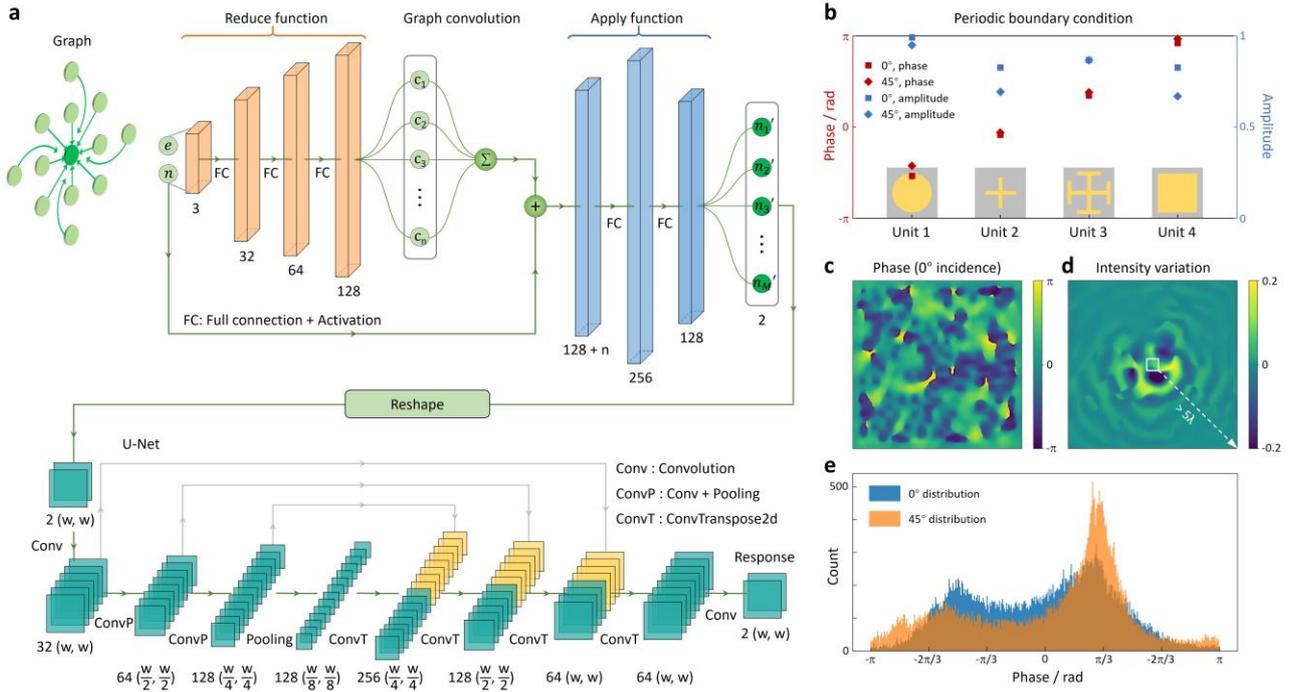

**Fig. 2 | Characterization of the long-range coupling effects based on GNN. a**, Framework of the GNN. The GNN is composed of a graph convolution module and a U-Net module. The graph convolutional module calculates the coupling effects through the reduce function and the apply function, while the U-Net module further processes the data to obtain the global response of the metasurface. **b**, Responses of four types of unit cells under periodic boundary condition. At incidence angles of 0° and 45°, all the units exhibit angle-insensitive characteristics. **c**, The phase response of a randomly generated 26 × 26 metasurface predicted by GNN at 0° incidence. **d**, Near-field intensity variation of the metasurface after modifying the central unit cells. When the unit cells within the white box are modified, the near-field over five wavelengths is affected due to the long-range coupling effects. **e**, Histograms of phase distributions of the metasurface predicted by GNN under 0° and 45° incidence. Both distributions cover the full 2π range but differ significantly, indicating incidence-angle-dependent response.

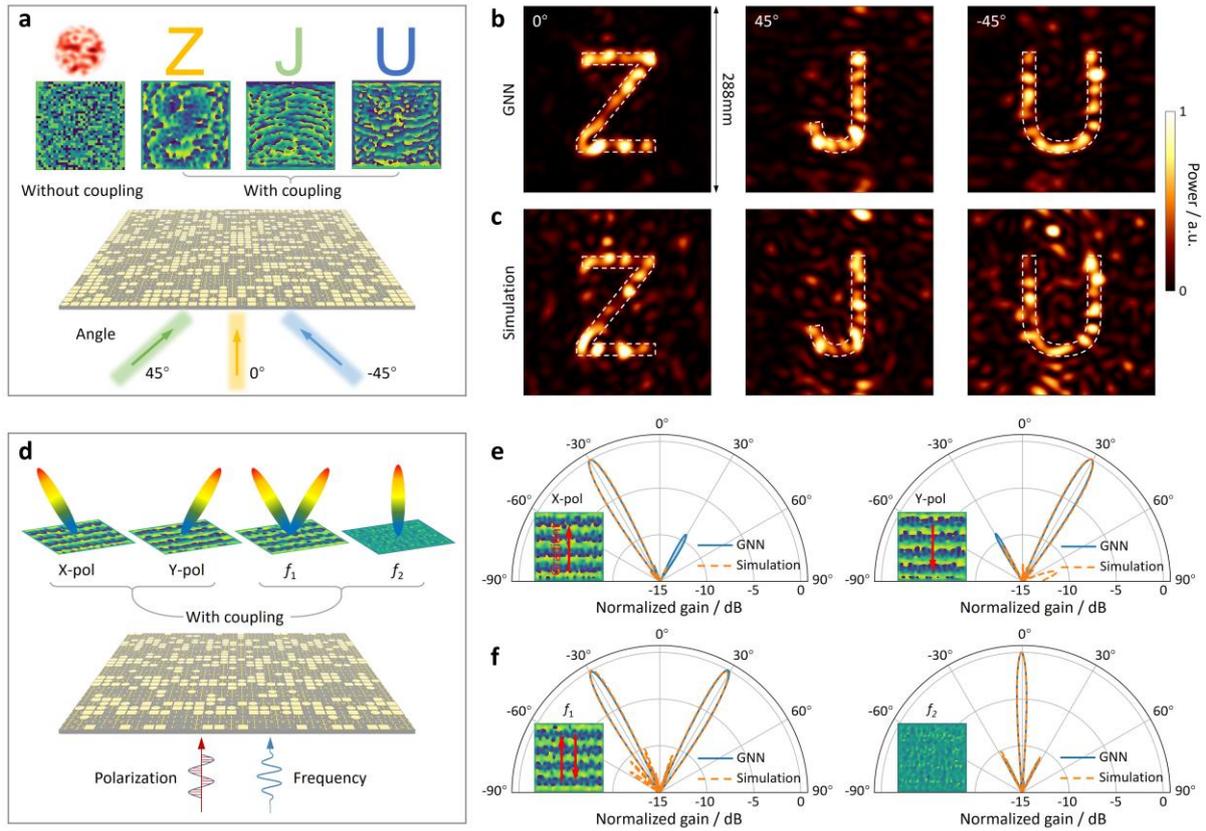

**Fig. 3 | Multi-channel metasurface multiplexing via long-range coupling. a**, Angular multiplexing mechanism. With long-range coupling, different incidence angles induce distinct phase profiles, thereby enabling unique holograms. Conversely, when coupling is neglected, the metasurface exhibits identical responses across all angles, making angular multiplexing unachievable. **b, c,** Holographic patterns generated by the metasurface at three incidence angles, as predicted by the GNN (**b**) and validated through full-wave simulation (**c**). **d**, Polarization and frequency multiplexing through coupling. Variations in incident polarization or frequency produce different far-field patterns, mediated by long-range coupling. **e**, Far-field results of polarization multiplexing. When illuminated by X-polarized or Y-polarized waves, the metasurface generates opposite phase gradients, resulting in distinct beam deflection angles. **f**, Far-field results of frequency multiplexing. The metasurface exhibits dual phase gradients at 11 GHz ($f_1$), and a nearly uniform phase at 12 GHz ($f_2$), leading to distinct beam shapes.

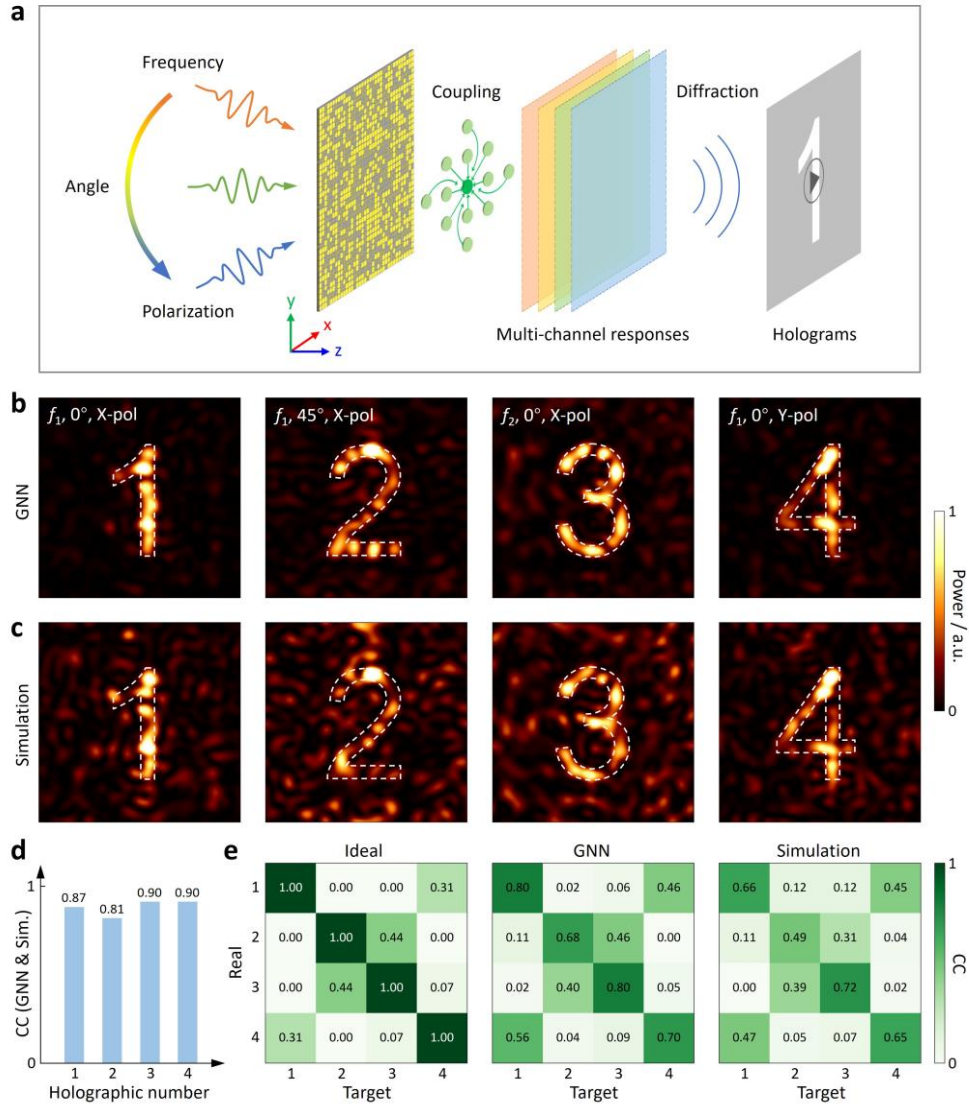

**Fig. 4 | Dynamic holograms of multi-dimensional multiplexing metasurface. a**, Dynamic holograms generated by multi-dimensional multiplexing. Variations in the incident wave's angle, polarization, and frequency change the coupling effects in the metasurface, inducing distinct holographic channels. **b**, Holographic patterns predicted by the GNN under different incident conditions. **c**, Corresponding patterns obtained from full-wave simulations. **d**, Correlation coefficients between the GNN-predicted and simulated holograms. **e**, Confusion matrices evaluating the channel isolation for the ideal, GNN-predicted, and simulated results.

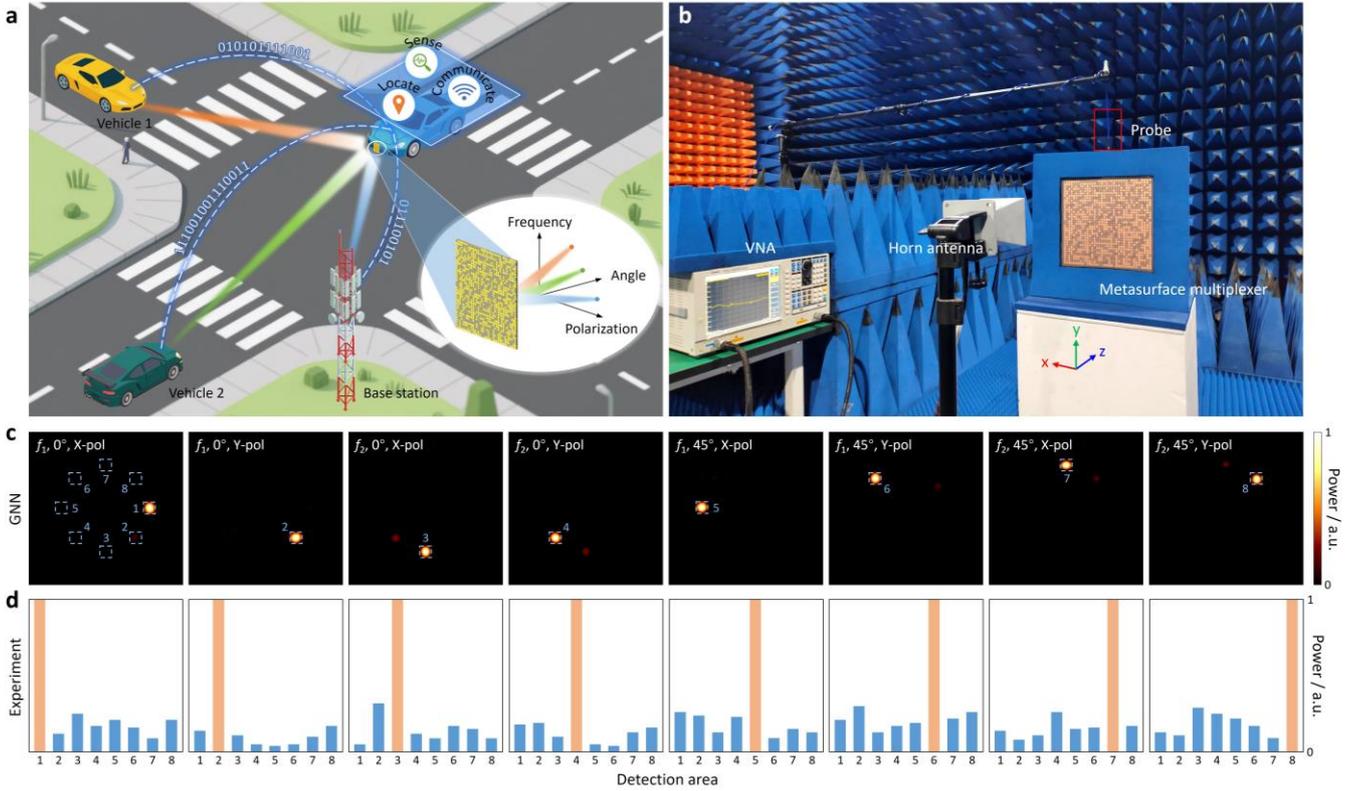

**Fig. 5 | Experimental demonstration of an 8-channel metasurface multiplexer. a**, Application scenario in vehicular networks. The compact and lightweight metasurface distinguishes incoming waves by their physical characteristics, thereby enabling integrated sensing, localization, and high-capacity communication. **b**, Experimental setup. The incidence angle is controlled by adjusting the horn antenna's rotation and elevation. **c**, GNN-predicted power distributions across the focal plane for different incident conditions. **d**, Experimentally measured power distributions at various areas under different incidence. The orange and blue bars denote the power in the target and non-target areas respectively, under various incidence conditions.